# Using Extreme Value Theory for Determining the Probability of Carrington-Like Solar Flares

## S. Elvidge and M. J. Angling

Space Environment and Radio Engineering Group, University of Birmingham, UK.

Corresponding author: Sean Elvidge (s.elvidge@bham.ac.uk)

**Key Points:**

- Extreme value theory is applied to GOES X-ray solar flare data
- Worst case of time scale of Carrington-like solar flare is shown to be 30 years
- Worst case of the largest flare in a 150-year period is shown to be an X90 flare

**Abstract**

By their very nature, extreme space weather events occur rarely and, therefore, statistical methods are required to determine the probability of their occurrence. Space weather events can be characterised by a number of natural phenomena such as X-ray (solar) flares, solar energetic particle (SEP) fluxes, coronal mass ejections and various geophysical indices (Dst, Kp, F10.7). In this paper extreme value theory (EVT) is used to investigate the probability of extreme solar flares. Previous work has assumed that the distribution of solar flares follows a power law. However such an approach can lead to a poor estimation of the return times of such events due to uncertainties in the tails of the probability distribution function. Using EVT and GOES X-ray flux data it is shown that the expected 150-year return level is approximately an X60 flare whilst a Carrington-like flare is a one in a 100-year event. It is also shown that the EVT results are consistent with flare data from the Kepler space telescope mission.

## 1 Introduction

In the popular press, solar flares have become synonymous with space weather events even though they do not directly cause most of the major space weather effects [*Hapgood*, 2012]. However, the magnitude of a flare usually provides an indication of the total amount of energy in a space weather event. Flare peak fluxes are classified by the letters, A, B, C, M and X where each letter indicates a flare one order of magnitude larger than the one preceding. An A1 flare has a peak X-ray flux of $1 \times 10^{-8}$ Wm$^{-2}$ (measured in the 0.1 to 0.8 nm range) whilst an X1 flare has $1 \times 10^{-4}$ Wm$^{-2}$. Extreme space weather events (solar superstorms) are often compared to the Carrington event of 1859 [*Carrington*, 1859]. The Carrington event is thought to be the largest observed space weather event in the last 200 years. The flare associated with the Carrington event has been estimated to be an X45 ± 5 (i.e. $45 \pm 5 \times 10^{-4}$ Wm$^{-2}$) [*Cliver and Dietrich*, 2013].

The tail of the distribution of solar flares has long been assumed to follow a simple power law [*Lu and Hamilton*, 1991; *Riley*, 2012], ${dN}/{dE} \approx E^{-\alpha}$ where $E$ is the flare energy, $N$ is the number of flares and α is the shape parameter [*Hudson*, 2010]. The value of α has been estimated to be between 1.7 and 2 [*Boffetta et al.*, 1999; *Aschwanden and Freeland*, 2012]. It should be noted that a power law distribution with shape parameter less than 2 has an undefined mean and standard deviation due to the distribution being "heavy-tailed". This means that all values are expected to occur eventually. Previous work on space weather risk has mostly been based on the assumption that the flare distribution follows such a law [*Riley*, 2012; *Cannon*, 2013]. An alternative approach by *Love* [2012] used Poisson-event occurrence rates to estimate the probability of another Carrington event. This paper uses of extreme value theory (EVT) to reanalyze the distribution of solar flares with the aim of estimating, with confidence intervals, space weather risk.

## 2 Extreme Value Theory

Extreme value theory (EVT) provides advanced tools for estimating probability distribution functions. Such an approach avoids any starting assumption about the underlying distribution [Coles, 2001]. EVT has a wide range of applications; for example in modelling metal alloy strengths [Tyron and Cruse, 2000], estimating extreme wind speeds [Della-Marta and et al., 2009] and a variety of uses in quantitative finance [Rocco, 2014]. In the various branches of space weather, EVT has been used to investigate the distribution of the daily Aa index (a measure of the disturbance of the Earth's magnetic field) [Silbergleit, 1999], disturbance storm time (Dst) index (an indication of the strength of the equatorial electrojet) [Tsubouchi and Omura, 2007], geomagnetic data (${dH}/{dt}$ where H is the horizontal geomagnetic component) [Thomson et al., 2011] and relativistic electron fluxes [Meredith et al., 2015].

EVT is mainly based around two theorems: the Fisher-Tippett-Gnedenko (FTG) theorem [*Coles*, 2001] and the Pickands-Balkema-de Haan (PBdH) theorem [*Balkema and de Haan*, 1974; *Pickands*, 1975]. FTG states that for a suitably normalized sample, an independent and identically distributed (*iid*) random variable converges to one of only three possible distributions: the Gumbel distribution [*Gumbel*, 1935], the Fréchet distribution [*Fréchet*, 1927] or the Weibull distribution [*Weibull*, 1951]. The distributions each have a shape (ξ), location (μ) and scale (σ) parameter which defines the distribution. However the requirement that the variable must be *iid* leads to difficulties in using EVT with raw data. Usually the data is not independent, however it can be made so by declustering (see Section 3).

The PBdH theorem (also known as the second theorem of EVT) defines an approach for modelling only the tail of an unknown distribution above a threshold value. The advantage of the PBdH is that it avoids modelling the entire distribution. Specifically, the theorem states that, for a large enough threshold value ($u$) the distribution of exceedances of this threshold, given that the random variable is greater than the threshold, is described by a generalised Pareto distribution (GPD) [*Leadbetter et al.*, 1983; *Coles*, 2001]:

$$H(y) = 1 - \left(1 + \frac{\xi y}{\sigma + \xi(u-\mu)}\right)^{\frac{-1}{\xi}}. \tag{1}$$

Where ξ is the shape parameter, μ the location, σ the scale and $u$ the threshold value. Determining the correct threshold value is crucial to the success of the PBdH theorem [*Coles*, 2001]. The value of $u$ should be one such that for all values greater than $u$, $u_0$, the GPD parameters associated with the exceedances over $u_0$ are the same (subject to a change of scale). Equivalently, for values of $u_0$ (which are greater than $u$,) the expectation of the exceedances of a random variable $X$, given that $X$ is greater than the threshold, $E(X - u_0 \mid X > u)$, should be a linear function of $u$. PBdH further states that the GPD variables are directly associated to the corresponding distribution from the FTG theorem. In particular the shape parameter ξ is equal in the two theorems. The parameters of the GPD are estimated using a maximum log-likelihood method.

## 3 Data

The solar X-ray flux data used in this study is from the X-ray Sensor (XRS) [*Machol and Viereck*, 2015] on the Space Environment Monitor subsystem of the NASA/NOAA Geostationary Operational Environmental Satellites (GOES) satellite missions. The raw data is collected approximately every three seconds and the data used in this work are one-minute averages of this raw data. There have been a number of GOES satellites since GOES-1 launched in October 1975. The main change in the data set arises from the switch from spinning satellites (GOES-7 and previous) to three-axis stabilized (GOES-8 onwards). A scaling factor is required to make the data from each GOES satellite consistent. To get the true X-ray flux value from the latest GOES satellites the data needs to be divided by 0.7. This ensures that the classification level (e.g. X5) is consistent across all the GOES satellites [Machol and Viereck, 2015]. Data from GOES satellites between 1986 and 2016 has been used in this study.

The XRS has also been shown to saturate during the most extreme events. During the storm of October/November 2003 the instrument saturated at a value of $17 \times 10^{-4}$ Wm$^{-2}$ (X17 flare), whereas the largest flare of this period is estimated to be X35 ($\pm$5) [Cliver and Dietrich, 2013]. More recent GOES satellites have not yet experienced such a large flare, but it is expected they will saturate at similar flux levels. The GOES dataset contains 11 saturation events; however, it is believed that one of these X17 flares is not a saturation event, but a true X17 event. It has been independently estimated that the flare on October 28th 2003 was an X17.2 flare [Blagoveshchensky et al., 2006]. Consequently, in this analysis, all but one X17 flare has been removed from the dataset.

The flare data is neither independent nor identically distributed. Flare events are usually recorded multiple times in the GOES data (since the fluxes remain high for longer than one minute). Thus a large flux value is often followed by other large fluxes. Also, the absolute number of events is related to the 11-year solar cycle: during solar maximum there is an increased number of flare events, and there is a corresponding decrease during solar minimum. Thus the underlying flare distribution varies temporally. One approach to dealing with these temporal variations would be to take the ten-year maximum value (thereby accounting for the solar cycle variation). However this is not currently practical since there is only data for approximately four solar cycles. Four data points is clearly not enough to do any analysis on, as such a different approach is required. The usual approach is to decluster the data [*Coles*, 2001]. The classic declustering process discounts contiguous data above a threshold, this is since it is assumed that contiguous events are actually a single event. It is assumed that the remaining data is then independent. For the GOES data the threshold flux value is set to that of a class X1 flare. For a new event to be counted there must have been 15 consecutive values (since 1 minute averaged data is used, this equates to 15 minutes) of X-ray fluxes less than $1 \times 10^{-4}$ Wm$^{-2}$ (M-class flux values or lower) between any two X flares. If a flux of greater than $1 \times 10^{-4}$ Wm$^{-2}$ is detected within the subsequent 15 minutes then the counter is reset to zero, and the largest flux is counted. It is conceivable that two consecutive X flares could come from different active regions and therefore be independent events. Ideally these would both be counted, but in this approach only one would be. Such analysis has not been possible given the lack of data from which active region each flare originated. However it is highly unlikely for two X class flares to occur in different regions within such a short time period, so missing data is not suspected to have a large impact on the results.

'One way to argue if the data set is independent or not, is to consider the autocorrelation. The autocorrelation with lag one for the original X-ray flare dataset is ~0.98. Figure 1 shows the autocorrelation value as a function of the number of consecutive minutes of fluxes less than $1 \times 10^{-4}$ Wm$^{-2}$ used in the declustering. Applying the declustering approach described in this Section drops the correlation value as the declustering gap increases. This continues up to a value of 15 after which the autocorrelation randomly fluctuates. The correlation, with a declustering gap of 15 minutes, drops to ~0.23. This provides confidence that the declustered flare dataset is independent and is suitable for use in EVT.

## 4 Results

The mean exceedance plot for the GOES X-ray flux data is shown in Figure 2. This shows some evidence of linearity for values of $u$ greater than $3.5 \times 10^{-4}$ Wm$^{-2}$ (and less than $12 \times 10^{-4}$ Wm$^{-2}$ where values of $u$ become unreliable due to the lack of data [*Coles*, 2001]). Therefore a value of $u = 3.5 \times 10^{-4}$ Wm$^{-2}$ is used as the GPD threshold value. An X3.5 flare is between 'Strong' and 'Severe' on the Space Weather Prediction Centre (SWPC) scales [*NOAA*, 2011], and using this value as the 'extreme' event boundary is a good compromise between rarity of the event, whilst having enough data for analysis. This choice results in a set of 171 exceedance values (i.e 171 flares greater than X3.5 in the dataset). Maximum log likelihood can then be used to estimate the GPD parameters, resulting in: $\hat{\sigma} = 2.98 \ (\pm 0.02) \times 10^{-4}$ and $\hat{\xi} = 0.26 \ (\pm 0.09)$, where the standard errors are found from the covariance matrix. Since $\xi > 0$ this is the Fréchet distribution [*Fréchet*, 1927], which has no upper limit.

One method for verifying the quality of the estimation of the distribution is to investigate the probability plot. For a set of $k$ threshold excesses, $y_{(1)} \leq \cdots \leq y_{(k)}$, for a given model $H$ the probability plot consists of the points:

$$\left( \frac{i}{k+1}, \ H\left(y_{(i)}\right); i = 1, \ldots, k \right), \tag{2}$$

where

$$H(y) = 1 - \left( 1 + \frac{\hat{\xi} y}{\hat{\sigma}} \right)^{-\frac{1}{\hat{\xi}}}. \tag{3}$$

where $\hat{\xi}$ and $\hat{\sigma}$ were found in the analysis. Such points should be roughly linear to give confidence that the model is a good estimate of the empirical distribution. Figure 3 shows the probability plot for the GOES X-ray flux data and the EVT model fit. It can be seen that the points are linear, and closely follow the unit diagonal. This gives confidence in the fit of the model.

Given the confidence in the estimated probability distribution the EVT fit can be used to estimate the largest expected flare in a given time period. For example, what is the largest flare we would expect to see in a 50-year period? To calculate this, the $m$-observation return level is given by [*Coles*, 2001]

$$x_m = u + \frac{\sigma}{\xi} \left[ \left( \frac{m d n_c}{n} \right)^{\xi} - 1 \right], \tag{4}$$

where $m$ is the year return level, $u$ the threshold value ($3.5 \times 10^{-4}$ Wm$^{-2}$), $\sigma$ ($2.98 \ (\pm 0.02) \times 10^{-4}$) and $\xi$ ($0.26 \ (\pm 0.09)$) the estimated GPD parameters, $d$ the number of observations in a year (525,600), $n_c$ the number of exceedances greater than $u$ (171) and $n$ the total number of observations (15,768,000). The return levels, and 95% confidence intervals, can then be produced. Figure 4 shows the X-ray flare return level plot.

Evidence of the validity of these returns can be found by considering data which is not included in the GPD fit. For example the X35 (±5) flare associated with the so-called "Halloween storm" on November 4th, 2003 [*Cliver and Dietrich*, 2013] which is included in Figure 4. It can be seen that this falls within the 95% confidence intervals of the EVT estimated returns.

Furthermore, Maehara *et al.* [2015] looked for superflares on Sun like stars using *Kepler* data between April 2009 and May 2013. The study found 1547 stars similar to our Sun (rotational period longer than 25 days, surface temperature between 5,300 K and 6,300 K and surface gravity of $4.0 < \log(g) < 4.8$). Amongst these stars they found evidence of 187 flares which had energies between $2 \times 10^{32}$ and $8 \times 10^{35}$ erg. These observations led Maehara *et al.* to the conclusion that X100 flares (~$10^{33}$ erg) could be expected on the Sun once in approximately 500-600 years. Such a result is within the 95% confidence intervals of our EVT analysis (Figure 4) giving further assurances to this method. It is interesting to note, that Aulanier *et al.* [2013] suggest that the largest flare possible on our Sun is ~X200.

Using these return levels it is estimated that a Carrington like flare (X45) is expected once in a ~100 year period, with 95% confidence intervals of 30 to 900 years. The extremely large confidence intervals highlight the difficulties in modelling extreme values with very little input data. Previous work has put the expected return time, with 95% confidence intervals, of a Carrington event at 79 years [2, 300] [*Riley*, 2012] and 159 years [4, ∞) [*Love*, 2012]. These estimates using EVT are broadly consistent with previous estimates. The largest possible flare of X200 (as suggested by Aulanier *et al.* [2013]) has a return time of 15,000 years (95% confidence intervals of 2,000 – 750,000 years).

## 5 Conclusions

Extreme value theory is used to provide rigorous statistical estimation of rare events. Part of that rigour is to acknowledge the errors in estimation, hence the large confidence intervals where there is no data. This paper has shown that using EVT on solar flare data results in an occurrence distribution that is consistent with both GOES X-ray flux data and data from the *Kepler* mission.

There are two ways for a national risk register to consider the 'worst case scenario': the return time of events of a particular intensity, or the largest expected event in a given time scale. If considering the worst case in terms of intensity of event, e.g. the return period of an X45 (Carrington) flare, then our analysis shows that the worst case should be considered as a one in 30 year event (the lower bound of the 95% confidence interval). A "reasonable" worst case would be to consider X45 as a one in 100 year event. If one considers the largest event in a given time scale, e.g. 150 years, then the worst case is an X90 flare and a "reasonable" worst case is an X60 flare.

**Acknowledgments**

The GOES data was collected from the NGDC website at http://ngdc.noaa.gov/stp/satellite/goes/.

# References

Aschwanden, M. J., and S. L. Freeland (2012), Automated solar flare statistics in soft X-rays over 37 years of GOES observations: the invariance of self-organized criticality during three solar cycles, *Astrophys. J.*, *754*(2).

Aulanier, G., P. Demoulin, C. J. Schrijver, M. Janvier, E. Pariat, and B. Schmieder (2013), The standard flare model in three dimensions II. Upper limit on solar flare energy, *Astron. Astrophys.*, *549*(A66).

Balkema, A., and L. de Haan (1974), Residual life time at great age, *Ann. Probab.*, *2*, 792–804.

Blagoveshchensky, D. V, J. W. MacDougall, and A. V Piatkova (2006), Ionospheric effects preceding the October 2003 Halloween storm, *J. Atmos. Solar-Terrestrial Phys.*, *68*(7), 821–831.

Boffetta, G., V. Carbone, P. Giuliani, P. Veltri, and A. Vulpiani (1999), Power Laws in Solar Flares:Self-Organized Criticality or Turbulence?, *Phys. Rev. Lett.*, *83*(22), 4662–4665.

Cannon, P. S. (2013), *Extreme space weather: impacts on engineered systems and infrastructure*.

Carrington, R. C. (1859), Description of a Singular Appearance seen in the Sun on September 1, 1859, *Mon. Not. R. Astron. Soc.*, *20*, 13–15.

Cliver, E. W., and W. F. Dietrich (2013), The 1859 space weather event revisited: limits of extreme activity, *Sp. Weather Sp. Clim.*, *3*(A31).

Coles, S. (2001), *An Introduction to Statistical Modeling of Extreme Values*, Springer, London.

Della-Marta, P. M., and et al. (2009), The return period of wind storms over Europe, *Int. J. Climatol.*, *29*, 437–459.

Fréchet, M. (1927), Sur la loi de probabilité de l'écart maximum, *Ann. Soc. Pol. Math.*, *6*.

Gumbel, E. J. (1935), Les valeurs extrêmes des distributions statistiques, *Ann. l'Institut Henri Poincaré*, *5*, 115–158.

Hapgood, M. (2012), Astrophysics: Prepare for the coming space weather storm, *Nature*, *484*, 311–313.

Hudson, H. S. (2010), Solar physics: Solar flares add up, *Nat. Phys.*, *6*, 637–638.

Leadbetter, M. R., G. Lindgren, and H. Rootzen (1983), *Extremes and Related Properties of Random Sequences and Processes*, Springer Verlag, New York.

Love, J. J. (2012), Credible occurrence probabilites for extreme geophysical events: Earthquakes, volcanic eruptions, magenetic storms, *Geophys. Res. Lett.*, *39*(L10301), doi:10.1029/2012GL051431.

Lu, E. T., and R. J. Hamilton (1991), Avalanches and the Distribution of Solar Flares, *Astrophys. J.*, *380*, L89–L92.

Machol, J., and R. Viereck (2015), *GOES X-ray Sensor (XRS) Measurements*.

Maehara, H., T. Shibayama, Y. Notsu, S. Notsu, S. Honda, D. Nogami, and K. Shibata (2015), Statistical properties of superflares on solar-type stars based on 1-min cadence data, *Earth, Planets Sp.*, *67*(59).

Meredith, N. P., R. B. Horne, J. D. Isles, and J. V Rodriguez (2015), Extreme relativistic electron fluxes at geosynchronous orbit: Analysis of GOES E > 2 MeV electrons, *Sp. Weather*, *13*(3), 170–184.

NOAA (2011), NOAA Space Weather Scales, Available from: http://www.swpc.noaa.gov/noaa-scales-explanation (Accessed 27 December 2016)

Pickands, J. (1975), Statistical inference using extreme order statistics, *Ann. Stat.*, *3*, 119–131.

Riley, P. (2012), On the probability of occurrence of extreme space weather events, *Sp. Weather*, *10*(S02012), doi:10.1029/2011SW000734.

Rocco, M. (2014), Extreme value theory in finance: a survey, *J. Econ. Surv.*, *28*, 82–108.

Silbergleit, V. M. (1999), The Most Geomagnetically Disturbed 24 Hours, *Stud. Geophys. Geod.*, *43*(2), 194–200.

Thomson, A. W. P., E. B. Dawson, and S. J. Reay (2011), Quantifying extreme behavior in geomagnetic activity, *Sp. Weather*, *9*(10).

Tsubouchi, K., and Y. Omura (2007), Long-term occurrence probabilities of intense geomagnetic storm events, *Sp. Weather*, *5*(12).

Tyron, R. G., and T. Cruse (2000), A. Probabilistic mesomechanics for high cycle fatigue life prediction, *J. Eng. Mater. Technol.*, *122*, 209–214.

Weibull, W. (1951), A statistical distribution function of wide applicability, *J. Appl. Mech. - ASME*, *3*, 293–297.

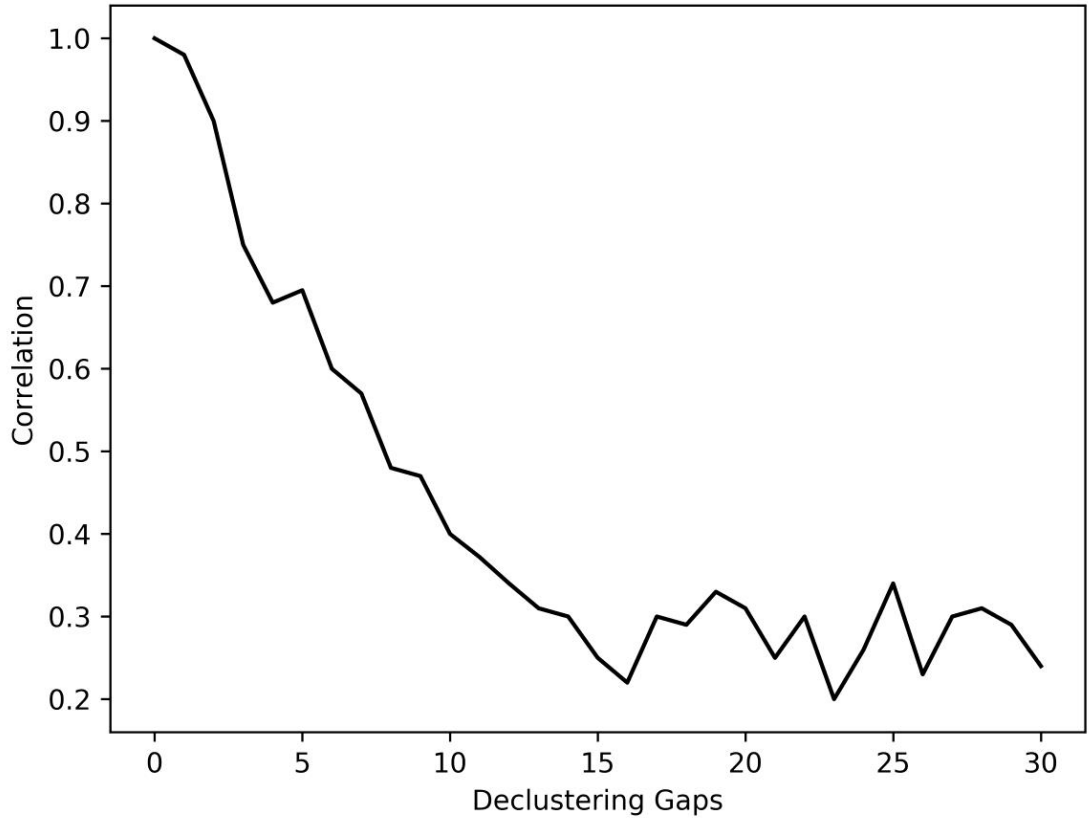

**Figure 1**. Autocorrelation with a lag of one for the X-ray flare data as a function of the declustering gap value (Section 3). The correlation continuously drops up to value of 15, after which the correlation randomly fluctuates.

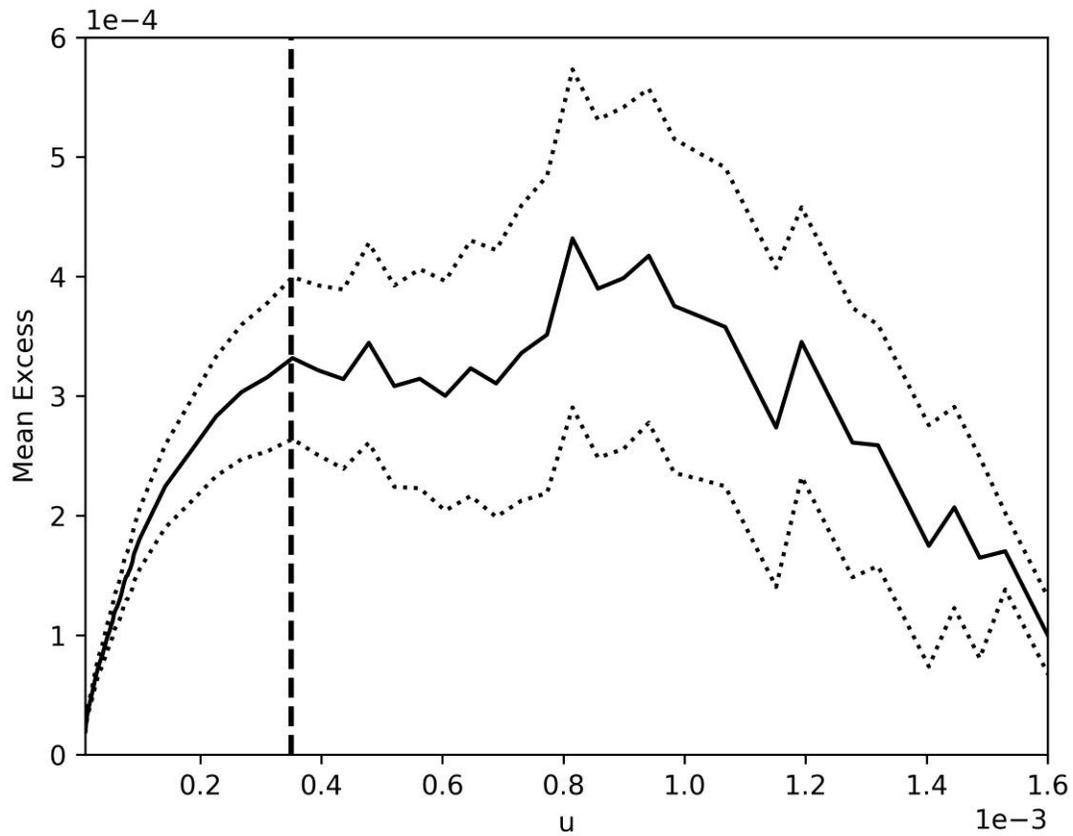

**Figure 2**. Mean exceedance plot for the GOES X-ray flare data. The x-axis represents increasing values of $u$, whilst the y-axis shows the mean exceedance for the given value. 95% confidence intervals are also plotted. The graph shows signs of linearity (between the confidence intervals) for values of $u$ greater than $3.5 \times 10^{-4}$ Wm$^{-2}$(X3.5 flares) and less than $12 \times 10^{-4}$ Wm$^{-2}$ where values of $u$ become unreliable due to the lack of data [Coles, 2001].

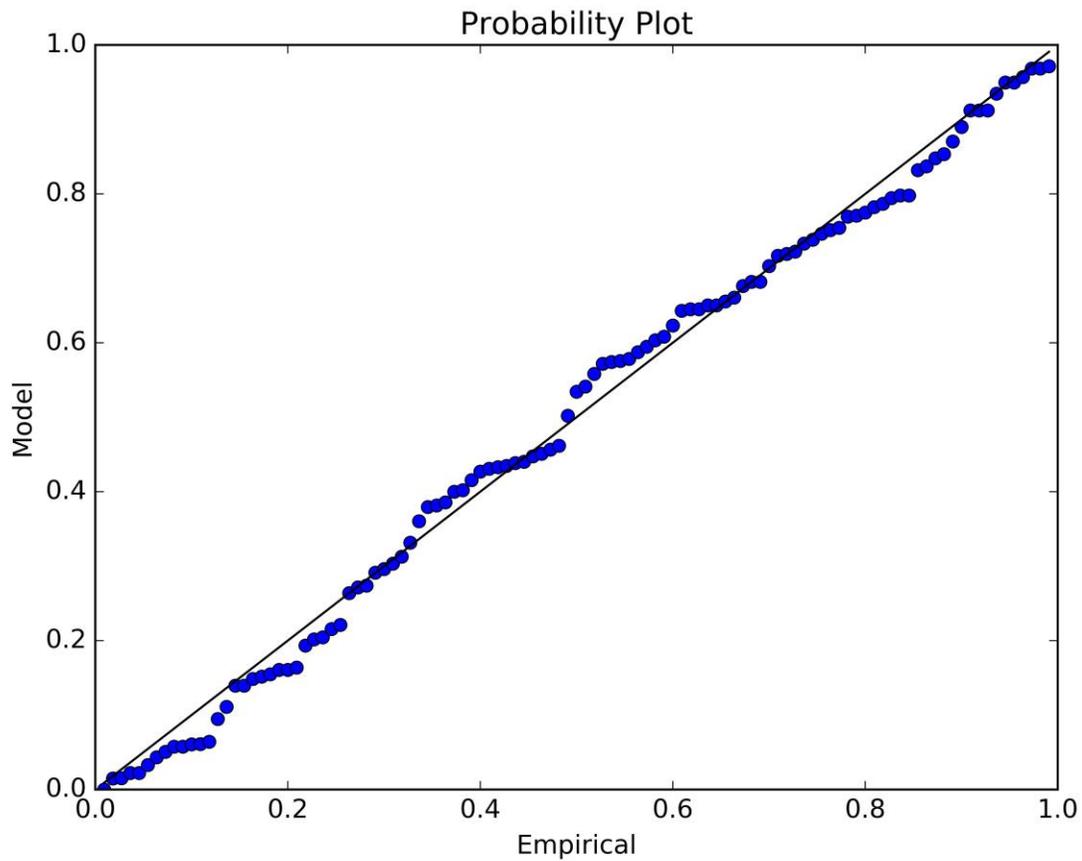

**Figure 3**. The probability plot compares the empirical and modelled distribution functions. These should be roughly linear (points close to the diagonal) for a good fit. It can be seen that there is a good fit between the X-ray flux data and the modelled EVT fit.

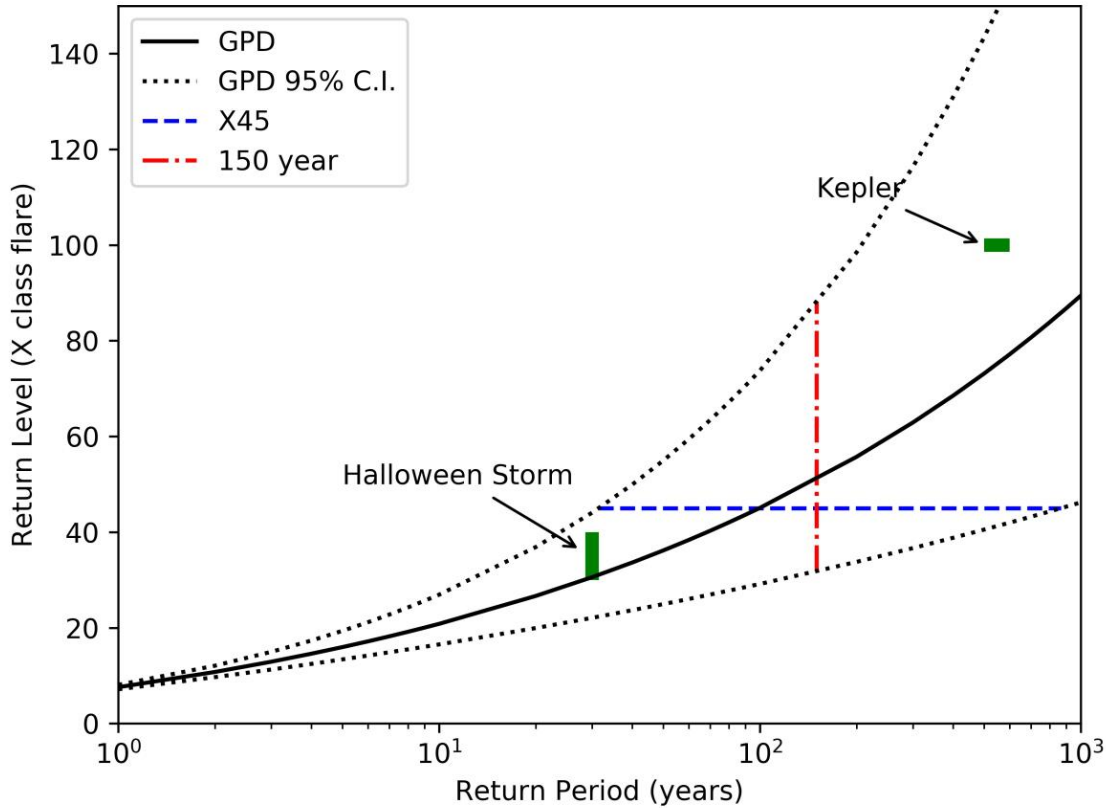

**Figure 4**. Return level plot, with 95% confidence intervals, for the GOES X-ray flare data. An X45 flare is expected once in a 100 year period, with 95% confidence intervals of 30 to 900 years. The largest expected flare in a 150 year period is an X57, with 95% confidence intervals of X30 to X90. Marked on the plot are the "Halloween Storm" from November 2003 which was reported to be an X35±5 flare, and results from the *Kepler* space telescope mission which report that an X100 flare is likely every 500-600 years [*Cliver and Dietrich*, 2013; *Maehara et al.*, 2015]. Neither results were included in the EVT analysis.